\documentclass[conference]{IEEEtran}
\IEEEoverridecommandlockouts

\usepackage{cite}
\usepackage{amsmath,amssymb,amsfonts}
\usepackage{algorithmic}
\usepackage{graphicx}
\usepackage{textcomp}
\usepackage{xcolor}
\usepackage{booktabs}
\usepackage{multirow}
\usepackage{graphicx}
\usepackage[hyphens]{url}
\usepackage{hyperref}
\usepackage[caption=false,font=footnotesize]{subfig}
\def\BibTeX{{\rm B\kern-.05em{\sc i\kern-.025em b}\kern-.08em
    T\kern-.1667em\lower.7ex\hbox{E}\kern-.125emX}}
\begin{document}

\title{SynSFX: Multi-Model Sound Effects Synthesis Dataset for Deepfake Detection and Evaluation\\}

\author{
\IEEEauthorblockN{
Linxi Li\IEEEauthorrefmark{1}\IEEEauthorrefmark{2}\textsuperscript{*},
Yuncong Yu\IEEEauthorrefmark{2}\textsuperscript{*},
Qianwei Guo\IEEEauthorrefmark{2},
Liwei Jin\IEEEauthorrefmark{2},
Yechen Wang\IEEEauthorrefmark{2},
Carsten Maple\IEEEauthorrefmark{1}
}
\IEEEauthorblockA{\IEEEauthorrefmark{1}University of Warwick, Coventry, United Kingdom}
\IEEEauthorblockA{\IEEEauthorrefmark{2}OfSpectrum, Inc., Los Angeles, CA, USA}
\thanks{\textsuperscript{*}Equal contribution.}
}
\maketitle

\begin{abstract}
While audio deepfake detection has advanced significantly, representative detectors show limited generalization to synthetic sound effects. Existing environmental audio datasets such as EnvSDD provide important initial resources, but remain limited in scale and generation provenance for studying isolated sound-effect deepfakes. To support this direction, we present SynSFX, a large-scale corpus of 43374 clips (26452 synthetic, 16922 real) spanning 7 popular text-to-audio models.
\end{abstract}


\section{Introduction}
Rapid advancements in deep generative modeling for audio---encompassing both speech and sound effects---have significantly outpaced the development of robust deepfake detection countermeasures \cite{khanjani}. State-of-the-art text-to-audio frameworks now synthesize high-fidelity acoustic events that are frequently perceptually indistinguishable from genuine recordings \cite{audiobox}. However, contemporary deepfake detection literature remains disproportionately anchored to speech synthesis and voice cloning paradigms \cite{asvspoof}. Consequently, a critical research gap persists regarding non-speech audio forensics, specifically encompassing environmental sounds, Foley, and ambient textures. Despite being equally susceptible to malicious manipulation as artificial speech, the targeted detection of synthetic sound effects remains a fundamentally underexplored domain.

However, sound effects differ fundamentally from speech \cite{Environmentalsurvey}. Unlike speech, which contains lexical and prosodic cues that detectors can exploit, non-speech audio such as sound effects and environmental sounds lacks structured linguistic information. Detection therefore relies purely on acoustic characteristics which modern generative models can closely imitate. As a result, systems trained primarily on speech data often fail to generalize to non-speech scenarios. 

This limitation is consequential. Sound effects are widely used in gaming, media production, live streaming, and safety-critical systems \cite{surveillance}. As generative models improve, the risk of misuse increases—from subtle manipulation in entertainment to injection of deceptive environmental sounds. To address this gap, we introduce SynSFX (Synthetic Sound Effects), a large-scale corpus for non-speech audio deepfake detection. The dataset includes diverse sound effects and environmental audio generated by multiple modern models, establishing a comprehensive benchmark beyond speech-focused evaluations. 

The primary contributions of this work are summarized as follows:
\begin{itemize}
    \item \textbf{A Large-Scale Benchmark with Diagnostic Control:} We release SynSFX, a comprehensive 178-hour corpus encompassing seven diverse text-to-audio architectures. Beyond providing massive scale and acoustic diversity, SynSFX uniquely features a specialized \textit{Shared Prompt Subset}. The shared-prompt subset provides a controlled resource for future prompt-matched analyses of generator-dependent artifacts.
    \item \textbf{Identification of Catastrophic Forgetting:} We empirically reveal that adapting speech-centric detectors exclusively to sound effects causes a catastrophic degradation in speech spoofing detection. We demonstrate that while joint-domain training mitigates this forgetting, critical generalization bottlenecks remain.
    \item \textbf{Exposing the Illusion of Generalization:} Through rigorous zero-shot evaluations on unseen generation architectures and feature space visualizations (t-SNE), we show that current models overfit to known synthesis artifacts rather than learning universal acoustic anomalies, establishing a baseline for future generalized audio forensics.
\end{itemize}

\section{Related Work}
\subsection{Speech Deepfake Detection}
Current research on audio deepfake detection has predominantly focused on human speech. Traditional and learning-based models, such as AASIST and RawNet2, have achieved remarkable success on standardized benchmarks like ASVspoof \cite{ASVspoof2019,ASVspoof2021} and FakeAVCeleb \cite{Khalid2021FakeAVCeleb}. Recent variants, including TO-RawNet \cite{Wang2023TORawnet} and twice-attention networks \cite{Yao2023TwiceAttention}, have further driven Equal Error Rates (EER) down to the 1-2\% range. However, as parallel datasets like WaveFake \cite{Frank2021WaveFake} and JMAD \cite{JMADUnspecified} reveal, speech-focused detectors often suffer from a persistent generalization gap under cross-dataset and cross-model transfer, indicating that robust detection remains challenging outside narrow, in-domain conditions.

\subsection{Advancements in Audio Generation Models}
The landscape of synthetic audio has recently expanded far beyond speech. Modern text-to-audio (TTA) and text-to-sound models—such as AudioCraft \cite{Copet2023MusicGen}, AudioLDM variants \cite{Liu2023AudioLDM, Liu2023AudioLDM2}, StableAudio \cite{StabilityAI2023StableAudio}, DiffSound \cite{Diffsound}, Make-An-Audio \cite{Huang2023MakeAnAudio}, MMAudio \cite{MMAUDIO}, and TangoFlux \cite{TANGOFLUX}—can now generate highly realistic ambient recordings, Foley, and general sound effects directly from text prompts. These modern generative models produce diverse acoustic patterns with significantly fewer spectral artifacts than earlier vocoder-based systems, rendering traditional artifact-driven detection methods increasingly obsolete and making deepfake detection exceptionally challenging.

\subsection{Deepfake Detection for General Audio and Sound Effects}
Despite the rapid evolution of general audio generation, the development of countermeasures for non-speech audio is bottlenecked by the lack of dedicated, structurally controlled datasets. Initial efforts, such as the Environmental Sound Deepfake Dataset (EnvSDD) \cite{EnvSDD}, pioneered the concept of environmental sound spoofing by providing a large-scale corpus of synthetic soundscapes. Concurrently, the recent CompSpoofV2 dataset \cite{Zhang2025CompSpoof} and the associated ESDD2 challenge \cite{ESDD2} significantly advanced the field by addressing \textit{component-level spoofing}—specifically, the complex scenario where synthetic environmental backgrounds are mixed with authentic foreground human speech. Consequently, state-of-the-art solutions like the EAT-AASIST framework \cite{EATAASIST} have achieved remarkable success in ESDD2 by focusing on separating and classifying these mixed components.

However, these foundational works leave a critical methodological gap in understanding the intrinsic artifacts of generative models. While CompSpoofV2 excels at evaluating mixed-source scenarios, its primary focus is on the artifacts introduced by the composition process, rather than the isolated generative flaws of the TTA models themselves. Conversely, while EnvSDD provides a relatively large volume of isolated synthetic audio, it relies on unconstrained, randomized text prompts across different architectures. This makes prompt-matched cross-generator analysis less direct. 

To address this gap, our work introduces \textbf{SynSFX}, a large-scale dataset focused on isolated deepfake sound effects. Comprising approximately 178 hours of diverse acoustic environments, SynSFX includes synthetic audio from seven TTA models with transparent generation provenance. In addition to generator-specific prompts, SynSFX contains a \textit{Shared Prompt Subset}, where the same text prompts are used across all generators. This subset enables prompt-matched comparisons of detector responses, helping distinguish semantic effects from generator-dependent variation. By focusing on isolated sound effects rather than mixed speech-background scenarios, SynSFX provides a benchmark for studying cross-generator robustness in non-speech audio forensics.

\section{The SynSFX Dataset}
SynSFX comprises a total of 43,374 audio clips, accumulating approximately 178 hours of high-fidelity recordings. The corpus is strategically designed to encompass both diverse authentic acoustic environments and state-of-the-art synthetic generations under standardized protocols. 

To systematically evaluate cross-model generalization, SynSFX incorporates synthetic outputs from seven prominent generative architectures. 

\subsection{Data Sources and Generative Models}
To ensure a rigorous and balanced evaluation, the dataset is composed of two primary partitions:
\begin{itemize}
    \item \textbf{Authentic Audio Subset:} To serve as the ground truth for deepfake detection, we curated 16,922 real audio clips from five established open-source repositories. The \textbf{AudioCaps} and \textbf{Clotho} \cite{Drossos2020Clotho} subsets provide natural environmental recordings paired with descriptive captions. \textbf{ESC-50} \cite{Drossos2020Clotho} contributes labeled environmental sounds across 50 everyday categories. \textbf{TACoS} \cite{Primus2025TACOS} adds activity-related audio events, while \textbf{WavCaps} \cite{Mei2024WavCaps} extends coverage to large-scale web-sourced ambient sounds.
    \item \textbf{Synthetic Audio Subset:} The fake counterparts were generated using seven state-of-the-art text-to-audio (TTA) models: \textbf{AudioLDM (v1/v2)} \cite{Liu2023AudioLDM}, \textbf{AudioCraft/AudioGen} \cite{Copet2023MusicGen}, \textbf{MMAudio} \cite{MMAUDIO}, \textbf{StableAudio} \cite{StabilityAI2023StableAudio}, \textbf{Make-An-Audio} \cite{Huang2023MakeAnAudio}, and \textbf{TangoFlux} \cite{TANGOFLUX}. These generators were strategically selected for their architectural diversity, encompassing diffusion-based, transformer-based, and latent generative approaches. In our experimental evaluations, these models are anonymized and denoted as A1--A7 (Table \ref{tab:SynSFX_sources}), respectively.
\end{itemize}

\subsection{Prompt Engineering and Expansion Pipeline}
To ensure immense acoustic diversity and mirror real-world user behaviors, SynSFX utilizes a structured prompt generation pipeline driven by advanced Large Language Models (LLMs), including ChatGPT and Gemini. The workflow initiates from concise baseline descriptions (e.g., ``footsteps on gravel''), which are subsequently expanded into contextually rich textual prompts (e.g., ``a person walking briskly on a gravel path under light rain''). This expansion mechanism introduces granular environmental variables and scene complexities, thereby enhancing the perceptual realism of the synthesized events. All expanded prompts undergo strict human filtration to eliminate semantic ambiguities or unsafe content.

\subsection{Corpus Architecture and Technical Specifications}
The prompt distribution within SynSFX utilizes a total of 28,350 unique textual prompts, mathematically structured to facilitate both comprehensive training and rigorous cross-model comparative analysis:
\begin{itemize}
    \item \textbf{Shared Prompt Subset:} Comprises 1,890 identical prompts provided universally to all evaluated TTA models. This subset serves as a controlled baseline, allowing researchers to isolate semantic variables and perform direct, cross-architecture comparisons of generation artifacts.
    \item \textbf{Exclusive Prompt Subsets:} The remaining distribution is evenly partitioned, with each generative model producing audio clips from unique prompts (detailed in Table \ref{tab:SynSFX_sources}) exclusive to that specific architecture, guaranteeing statistical balance.
\end{itemize}

To preserve inherent generation artifacts, all audio clips are stored in uncompressed WAV format using the native sample rate of their source. Diffusion-based generators (AudioCraft, AudioLDM variants, Make-An-Audio) produce 16 kHz audio, while MM-Audio, StableAudio, and TangoFlux output at 44.1 kHz. The real subsets are primarily recorded at 44.1 kHz, with the exception of AudioCaps at 22.0 kHz. For benchmarking, SynSFX is released with predefined train, validation, and test splits to ensure standardized evaluation across future research.

\section{Experiments \& Results}
\subsection{Baselines}
To establish a benchmark on SynSFX, we evaluated three distinct, representative deepfake detection architectures under a strict zero-shot protocol (i.e., evaluated directly on SynSFX without any domain-specific fine-tuning). The selected baselines include two well-established speech-centric models (\textbf{AASIST} and \textbf{RawNet2}) and a recent state-of-the-art framework capable of generalized audio processing (\textbf{EAT-AASIST}). The evaluation results are detailed in Table \ref{tab:baseline_compare}.

\subsection{Collapse of Speech-Centric Detectors}
As summarized in Table \ref{tab:baseline_compare}, both AASIST and RawNet2 exhibit severe performance degradation, performing near random chance levels: RawNet2 yields an Equal Error Rate (EER) of 49.94\%, while AASIST collapses entirely with an EER of 60.84\% and an exceptionally high False Acceptance Rate (FAR). 

This profound confusion pattern is mechanically predictable. Both AASIST and RawNet2 were originally optimized for the ASVspoof benchmarks, heavily relying on speech-intrinsic properties such as fundamental frequency subbands, vocal tract formants, and linguistic prosody. Because the SynSFX corpus fundamentally lacks these structured lexical and human-vocal cues, the traditional detectors fail to anchor onto any meaningful features, resulting in random feature-space distributions.

\subsection{Partial Success and Persistent Gaps of Generalized Model}
\label{partialsuccess}
While pure speech-trained models perform poorly, \textbf{EAT-AASIST} shows better zero-shot transfer on SynSFX, achieving an EER of \textbf{23.71\%} and an ROC AUC of \textbf{0.8590}. This relative robustness is likely due to its EAT backbone and prior training on ESDD2 \cite{ESDD2}, where the model is exposed to synthetic environmental components mixed with speech.

However, its performance remains below the in domain results, indicating that mixed-audio environmental spoofing and isolated sound-effect deepfake detection are not equivalent. These results motivate SynSFX as a dedicated benchmark for studying non-speech audio deepfake detection under isolated and multi-generator conditions.

\begin{table}[t]
\centering
\caption{Models and datasets used in constructing the SynSFX corpus. Durations are in hours. The shared prompt subset contains 1,890 prompts that are
generated by each AI generator; these clips are included in each model's total and are not counted as an additional generator.}
\vspace{1mm}
\begin{tabular*}{\columnwidth}{@{\extracolsep{\fill}} l r r r r @{}}
\hline
\textbf{Model / Dataset} & \textbf{Shared} & \textbf{Model-specific} & \textbf{\# Clips} & \textbf{Duration [h]} \\
\hline
A1 AudioCraft     & 1890 & 1890 & 3780 & 7.9  \\
A2 AudioLDM1      & 1890 & 1890 & 3780 & 7.8  \\
A3 AudioLDM2      & 1890 & 1890 & 3780 & 7.9  \\
A4 MMAudio        & 1890 & 1890 & 3780 & 6.8  \\
A5 Make-An-Audio  & 1890 & 1890 & 3780 & 11.3 \\
A6 Stable Audio   & 1890 & 1888 & 3778 & 10.5 \\
A7 TangoFlux      & 1890 & 1884 & 3774 & 8.9  \\
\hline
\textbf{Total Generated} & \textbf{13230} & \textbf{13222} & \textbf{26452} & \textbf{61.1} \\
\hline
AudioCaps        & -- & -- & 4000 & 11.0 \\
Clotho           & -- & -- & 3839 & 24.0 \\
ESC-50           & -- & -- & 2000 & 2.8  \\
TACoS            & -- & -- & 5000 & 31.1 \\
WavCaps          & -- & -- & 2083 & 50.9 \\
\hline
\textbf{Total Real} & -- & -- & \textbf{16922} & \textbf{119.7} \\
\hline
\end{tabular*}
\label{tab:SynSFX_sources}
\end{table}

\begin{table}[th]
\centering
\caption{Zero-shot Evaluation of Baseline Deepfake Detectors on SynSFX evaluation subset(Without Fine-tuning)}
\vspace{1mm}
\begin{tabular}{l c c c}
\hline
\textbf{Metric} & \textbf{AASIST} & \textbf{RawNet2} & \textbf{EAT-AASIST} \\ 
\hline
True Fake (TP) & 3698 & 4727 & 7205 \\
False Real (FN) & 5746 & 4717 & 2239 \\
False Fake (FP) & 10295 & 8451 & 4012 \\
True Real (TN) & 6627 & 8471 & 12910 \\
Equal Error Rate (EER) & 60.84\% & 49.94\% & \textbf{23.71\%} \\
Threshold at EER & 0.00210 & -0.02142 & 0.11014 \\
ROC AUC & 0.3609 & 0.5034 & \textbf{0.8498} \\
PR AUC & 0.2744 & 0.3902 & 0.7902 \\
F1 Score & 0.3158 & 0.4179 & 0.6974 \\
False Acceptance Rate (FAR) & 0.6082 & 0.4994 & 0.2371 \\
False Rejection Rate (FRR) & 0.6082 & 0.4995 & 0.2371 \\ 
\hline
\end{tabular}
\label{tab:baseline_compare}
\end{table}

\subsection{Finetuning Protocol}
To investigate whether domain-specific training can overcome the limitations observed in the zero-shot evaluations, we fine-tuned both AASIST and EAT-AASIST on the SynSFX corpus. Both architectures were optimized with binary supervision, classifying inputs as authentic or synthetic. To prevent data leakage, all clips were assigned globally unique identifiers, ensuring mutually exclusive training, validation, and evaluation splits.

To reduce preprocessing-related shortcuts, all audio was converted to mono, resampled to 16 kHz, and peak-normalized before training and evaluation. AASIST and EAT-AASIST used fixed input lengths of 64,600 and 64,000 samples, respectively; longer clips were cropped (random crop during training and center crop during validation/testing), while shorter clips were symmetrically zero-padded.

Both models were initialized from their official pre-trained weights \cite{EATAASIST,Jung2022AASIST} and fine-tuned end-to-end for 60 epochs on a single RTX 4090D GPU without architectural modification. We used AdamW with a learning rate of $1\times10^{-4}$, weight decay of $1\times10^{-5}$, and CosineAnnealingWarmRestarts ($T_0=10$, $T_{\mathrm{mult}}=2$). Batch sizes were 32 for AASIST and 16 for EAT-AASIST.

\subsection{Evaluation}
To systematically quantify both in-domain retention and cross-domain generalization, we established a rigorously stratified evaluation protocol consisting of two mutually exclusive test sets. 

\textbf{In-Domain (Seen Generators) Test Set:} To measure baseline fine-tuning efficacy, we extracted a dedicated test split directly from the SynSFX corpus. This in-domain evaluation set comprises \textbf{4,338} audio clips, totaling approximately 16 hours of acoustic data. While strictly excluded from training dataset at sample level, this subset exposes the detectors to the same acoustic domains, prompt structures, and text-to-audio architectures (A1–A7) encountered during the fine-tuning phase.

\textbf{Out-of-Domain (Unseen Generators) Test Set:} To rigorously assess cross-model transferability and zero-shot robustness, we constructed an entirely independent out-of-domain evaluation set comprising \textbf{1,113} audio clips. To guarantee a statistically sound evaluation, this set is symmetrically partitioned to maintain a strict 1:1 class balance:
\begin{itemize}
    \item \textit{Synthetic Subset (50\%):} Generated using a state-of-the-art, proprietary commercial text-to-audio API. By utilizing a closed-source architecture entirely absent from the SynSFX training phase, we ensure the detection models have zero prior exposure to its specific synthesis artifacts.
    \item \textit{Authentic Subset (50\%):} Sampled exclusively from the UrbanSound8K \cite{UrbanSound} dataset. This provides a distinct distribution of real-world environmental recordings and Foley sound effects that remain completely unrepresented in the primary training corpus.
\end{itemize}

By evaluating the fine-tuned models across these two distinct sets, we can accurately decouple genuine acoustic representation learning from mere overfitting to generator-specific flaws.

\subsection{In-Domain Efficacy and Catastrophic Forgetting}
As detailed in Table \ref{tab:comprehensive_results}, fine-tuning AASIST exclusively on the SynSFX corpus yields substantial performance improvements within the non-speech domain. The standard AASIST model successfully reduces its Equal Error Rate (EER) from the near-random 60.84\% (zero-shot) to an impressive 3.23\%. Similarly, the more advanced EAT-AASIST framework achieves a highly discriminative EER of 2.36\% with an F1 score of 0.9806. These in-domain results validate the utility of the SynSFX dataset, demonstrating that synthetic sound effects possess learnable, domain-specific generation artifacts that deep architectures can successfully extract when provided with sufficient domain-targeted supervision.

However, this dramatic improvement in sound effect forensics comes at a severe cost to the models' original capabilities. When the SynSFX-only fine-tuned AASIST model is evaluated back on its original speech deepfake testing set \cite{ASVspoof2019}, we observe a phenomenon of \textit{catastrophic forgetting}. The speech evaluation EER spikes to 33.61\% (Table \ref{tab:comprehensive_results}), indicating a profound degradation in voice spoofing detection. 

This functional collapse mechanically suggests that the network shifts its decision boundary to optimize for the heterogeneous spectral noise, diverse sound-effect spectra, and generator-specific flaws in synthetic effects, thereby "unlearning" the more consistent harmonic structures and vocal-tract cues required for authenticating human speech. This trade-off confirms that adapting a single-domain model via naive fine-tuning is inherently insufficient for generalized audio forensics, necessitating joint-domain training paradigms.

  \begin{table*}[t!]
  \centering
  \caption{Comprehensive cross-domain evaluation of fine-tuned detectors. \textit{SynSFX-Only} uses only sound-effect data, while \textit{SynSFX + Speech} uses joint-domain training. The unseen evaluation is further decomposed to separate fake-generator shift from real-domain shift.}
  \renewcommand{\arraystretch}{1.2}
  \begin{tabular}{ll l cccc}
  \toprule
  \textbf{Architecture} & \textbf{Training Paradigm} & \textbf{Evaluation Set} & \textbf{EER} & \textbf{ROC AUC} & \textbf{PR AUC} &
  \textbf{F1 Score} \\
  \midrule

  \multirow{10}{*}{\textbf{AASIST}}
  & \multirow{5}{*}{SynSFX-Only}
  & SynSFX Test (Seen)             & 3.23\%  & 0.9959 & 0.9975 & 0.9733 \\
  & & Speech (Out-of-Domain)       & 33.61\% & 0.7359 & 0.9626 & 0.7799 \\
  & & Seen Real + Unseen Fake      & 30.21\% & 0.7639 & 0.5529 & 0.5326 \\
  & & Unseen Real + Seen Fake      & 4.28\%  & 0.9955 & 0.9990 & 0.9738 \\
  & & Unseen Real + Unseen Fake    & 26.17\% & 0.7960 & 0.7817 & 0.7381 \\
  \cmidrule{2-7}
  & \multirow{5}{*}{SynSFX + Speech}
  & SynSFX Test (Seen)             & 3.76\%  & 0.9927 & 0.9941 & 0.9689 \\
  & & Speech (In-Domain)           & 3.61\%  & 0.9942 & 0.9993 & 0.9795 \\
  & & Seen Real + Unseen Fake      & 34.33\% & 0.7228 & 0.4675 & 0.4854 \\
  & & Unseen Real + Seen Fake      & 5.43\%  & 0.9878 & 0.9974 & 0.9663 \\
  & & Unseen Real + Unseen Fake    & 37.23\% & 0.6687 & 0.6377 & 0.6277 \\

  \midrule

  \multirow{11}{*}{\textbf{EAT-AASIST}}
  & Pre-trained Baseline
  & Speech (Source Domain)         & 4.05\%  & 0.9932 & 0.9992 & 0.9770 \\
  \cmidrule{2-7}
  & \multirow{5}{*}{SynSFX-Only}
  & SynSFX Test (Seen)             & \textbf{2.36\%} & \textbf{0.9978} & \textbf{0.9986} & 0.9806 \\
  & & Speech (Out-of-Domain)       & 24.04\% & 0.8425 & 0.9797 & 0.8500 \\
  & & Seen Real + Unseen Fake      & 25.63\% & 0.8222 & 0.6006 & 0.5887 \\
  & & Unseen Real + Seen Fake      & 2.16\%  & 0.9976 & 0.9995 & \textbf{0.9868} \\
  & & Unseen Real + Unseen Fake    & 30.13\% & 0.7660 & 0.7684 & 0.6990 \\
  \cmidrule{2-7}
  & \multirow{5}{*}{SynSFX + Speech}
  & SynSFX Test (Seen)             & 2.45\%  & \textbf{0.9978} & \textbf{0.9986} & 0.9798 \\
  & & Speech (In-Domain)           & 5.25\%  & 0.9891 & 0.9987 & 0.9700 \\
  & & Seen Real + Unseen Fake      & 27.50\% & 0.7989 & 0.5831 & 0.5652 \\
  & & Unseen Real + Seen Fake      & 2.16\%  & 0.9976 & 0.9995 & \textbf{0.9868} \\
  & & Unseen Real + Unseen Fake    & 32.19\% & 0.7497 & 0.7578 & 0.6781 \\

  \bottomrule
  \end{tabular}
  \label{tab:comprehensive_results}
  \end{table*}
\subsection{Mitigating Forgetting via Joint-Domain Training}
To address the severe degradation in speech forensics, we investigated a joint-domain training paradigm. The objective is to determine whether foundational acoustic models can maintain dual decision boundaries—preserving human vocal tract priors while simultaneously learning the heterogeneous artifacts of synthetic environmental sounds. 

To facilitate this, we integrated a large-scale speech deepfake corpus derived from the ASVspoof 2019 Logical Access (LA) dataset \cite{ASVspoof2019}. The joint fine-tuning partition incorporates 25,380 speech utterances (24.15 hours, comprising 22,800 synthetic and 2,580 authentic clips), while validation utilizes 24,844 utterances (24.00 hours, 22,296 synthetic and 2,548 authentic). For a rigorous speech-domain evaluation, we utilized an isolated evaluation split comprising 71,237 utterances (61.50 hours). This massive test set ensures the models are evaluated against diverse, unseen speech spoofing algorithms that were strictly excluded from the training phase.

As demonstrated in Table \ref{tab:comprehensive_results}, interleaving speech data into the fine-tuning process successfully rescues both architectures from catastrophic forgetting. For the standard AASIST model, joint-domain training restores the speech evaluation EER to an exceptional 3.61\% (a dramatic recovery from 33.61\%), while simultaneously maintaining a highly competitive EER of 3.76\% on the SynSFX test set. 

The EAT-AASIST framework exhibits an even stronger capacity for multi-domain representation. Under the joint-training paradigm, it achieves a 2.45\% EER and an F1 score of 0.9798 on synthetic sound effects, alongside a restored 5.25\% EER on speech. These results empirically confirm that joint-domain training acts as an effective stabilizer. It suggests that the network has the capacity to simultaneously map the acoustic signatures of both domains without mutual exclusion, even with minor domain-specific performance degradation.

However, while this joint paradigm successfully solves the issue of cross-domain memory retention, it introduces the final and most critical challenge of audio deepfake detection: zero-shot generalization to entirely unseen generators.

\subsection{The Illusion of Generalization: Overfitting to Generator Artifacts}
\label{illusion}
While joint-domain training mitigates cross-domain forgetting, Table~\ref{tab:comprehensive_results} shows that zero-shot generalization to unseen generators remains the dominant bottleneck. To separate fake-generator shift from real-domain shift, we evaluate two diagnostic settings: \textit{Seen Real + Unseen Fake} and \textit{Unseen Real + Seen Fake}. Across both architectures and training paradigms, performance degrades substantially when the fake samples come from the unseen generator. For example, joint-trained AASIST reaches 34.33\% EER on \textit{Seen Real + Unseen Fake}, while joint-trained EAT-AASIST obtains 27.50\% EER. In contrast, replacing the real side with UrbanSound8K while keeping seen SynSFX generators remains much less harmful: AASIST and EAT-AASIST achieve 5.43\% and 2.16\% EER, respectively, under \textit{Unseen Real + Seen Fake}.

This decomposition indicates that the main source of out-of-domain failure is the unseen synthetic generator rather than the authentic-source shift alone. When both shifts are combined in \textit{Unseen Real + Unseen Fake}, performance further degrades to 37.23\% EER for joint-trained AASIST and 32.19\% EER for joint-trained EAT-AASIST. These results suggest that current detectors still rely heavily on generator-specific artifacts learned from the seen SynSFX models, limiting their robustness to novel TTA architectures.

\begin{figure*}[t]
    \centering
    \subfloat[AASIST]{
        \includegraphics[width=0.45\textwidth]{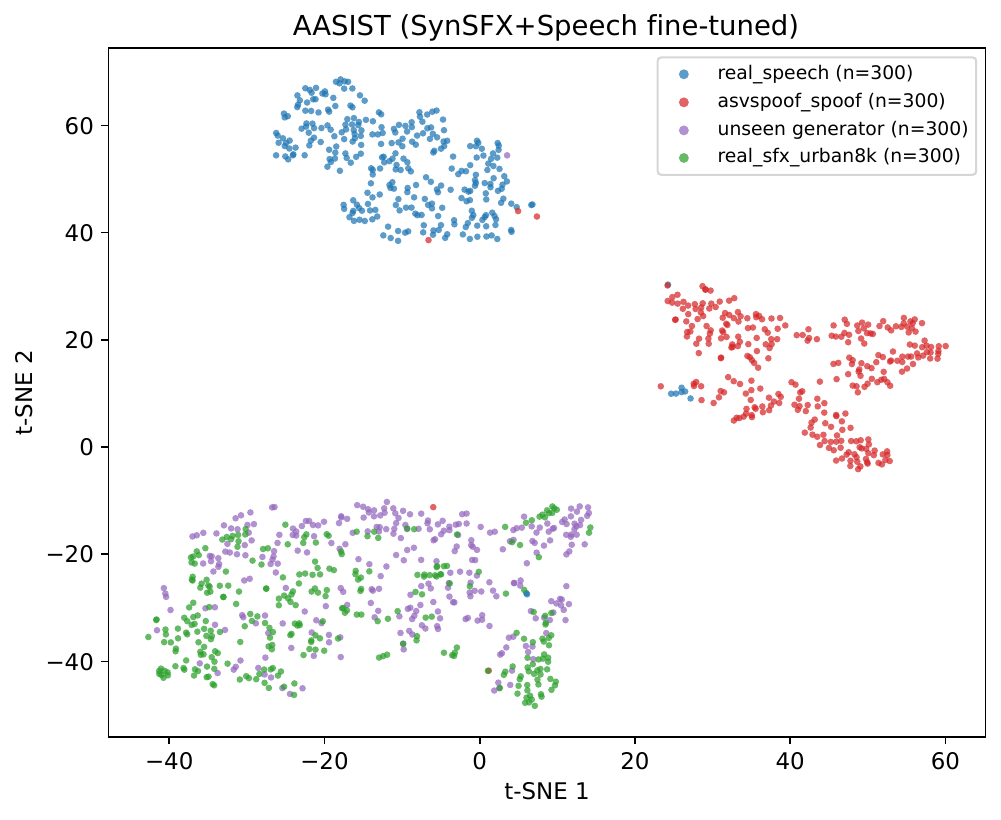}
        \label{fig:tsne_aasist}
    }
    \hfil 
    \subfloat[EAT-AASIST]{
        \includegraphics[width=0.45\textwidth]{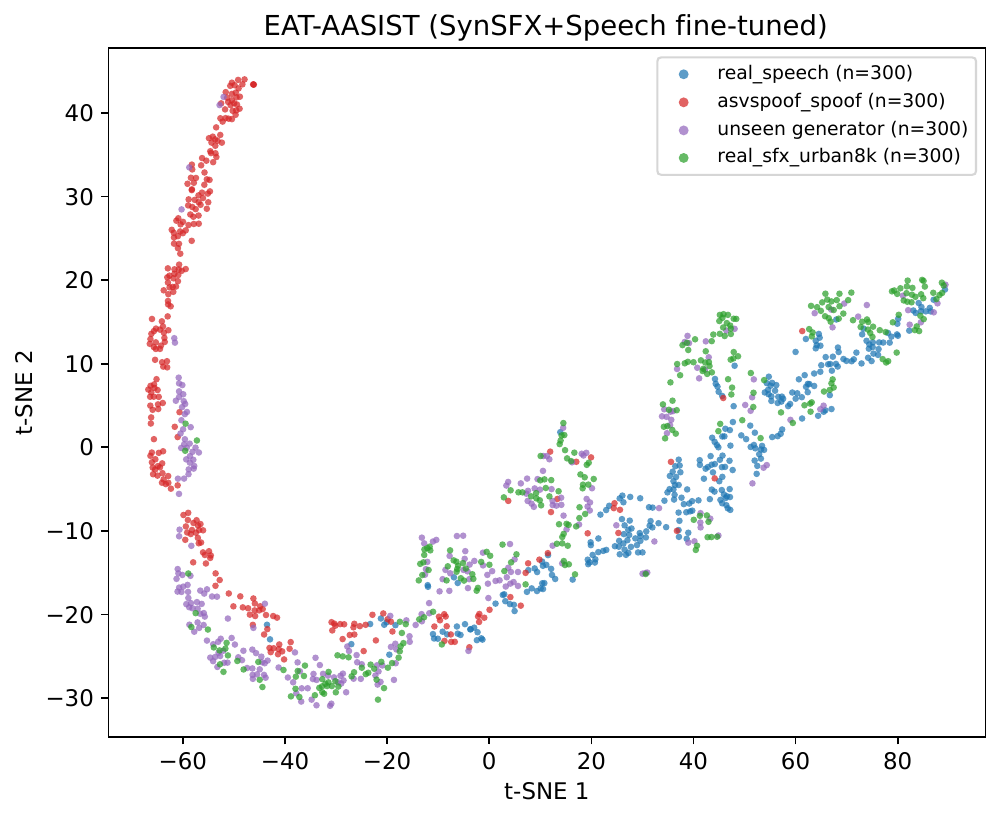}
        \label{fig:tsne_eat}
    }
    \caption{t-SNE visualization of the penultimate-layer embeddings for the joint-trained (a) AASIST and (b) EAT-AASIST models. Both architectures maintain linear separability for speech domains (red vs. blue) but suffer from severe feature entanglement when processing unseen non-speech audio (green vs. purple).}
    \label{fig:tsne}
\end{figure*}

\textbf{Feature Space Visualization:} To visually substantiate the challenge of general sound effect forensics, we projected the penultimate-layer embeddings of both joint-trained models into a 2D plane using t-SNE (Figure \ref{fig:tsne}). The projection contrasts four subsets: authentic speech, synthetic speech, real sound effects (UrbanSound8K), and unseen synthetic sound effects (A8).

The visualizations highlight a stark contrast between the speech and non-speech domains. While AASIST and EAT-AASIST exhibit different mapping behaviors—forming discrete clusters and a continuous U-shaped manifold, respectively—both models successfully establish clear linear separability between authentic and synthetic speech. Crucially, this discriminative structure entirely collapses for unseen non-speech audio. In both plots, real sound effects (green) and unseen synthetic effects (purple) are catastrophically entangled with no discernible decision boundary.

This performance degradation indicates that current deepfake detection paradigms suffer from an ``illusion of generalization.'' Mechanically, the models are not learning the fundamental, physics-based acoustic discrepancies between natural and synthesized environments. Instead, they are overfitting to the specific digital signatures, phase distortions, and diffusion noises inherent to the seven generative architectures (A1--A7) encountered during training. When confronted with a novel commercial architecture lacking these exact spectral fingerprints, the decision boundaries transfer poorly.

\textbf{Aligning with Broader Forensics Challenges:} This phenomenon is not unique to non-speech audio; it perfectly mirrors the well-documented generalization gaps historically observed in speech anti-spoofing. Foundational studies evaluating cross-corpus robustness (such as the ASVspoof 2021 Challenge \cite{ASVspoof2021} and cross-database evaluations on WaveFake \cite{Frank2021WaveFake}) have repeatedly demonstrated that models optimized on specific neural vocoders frequently degrade to near-random chance when tested on unseen generation algorithms. Our empirical results confirm that this ``overfitting to generator artifacts'' is equally, if not more, severe in the highly unstructured domain of general sound effects.

\textbf{Shared-Prompt Diagnostic Analysis}
Since EAT-AASIST already shows partial zero-shot discrimination on SynSFX before fine-tuning (Section~\ref{partialsuccess}), we use it to analyze the shared-prompt subset. For each shared prompt, we compute the variance and range of fake scores across the seven generators, and then average these prompt-level statistics over all shared prompts. Before adaptation, EAT-AASIST obtains an overall fake-score mean of 0.5135, its per-prompt range remains large (mean/median: 0.8915/0.9461), indicating that detector confidence varies strongly across generators even under identical semantic prompts.

After SynSFX fine-tuning, the overall fake-score mean increases to 0.9865, while the per-prompt variance mean drops from 0.1413 to 0.0082 and the range mean/median decreases to 0.0899/0.0037. This suggests that SynSFX supervision substantially improves cross-generator consistency for seen generators, although residual generator-specific or prompt-generator interaction effects may still influence detector confidence.

\textbf{The Foundational Value of SynSFX:} The cross-model collapse in Section \ref{illusion} illuminates the critical necessity of the SynSFX corpus. It suggests that naive fine-tuning isn't sufficient for robust audio forensics. To break this bottleneck, the community must transition toward learning generalized synthetic acoustic representations. 

SynSFX is strategically engineered to facilitate this exact paradigm shift through its \textbf{Shared Prompt Subset}. While different text-to-audio models interpret and render semantics with inherent stochastic variance, this shared subset provides an mechanism to control for macro-semantic variables. By ensuring that identical acoustic events (e.g., a ``footstep'') are represented across diverse generator distributions, it actively mitigates content-bias. This design encourages future methods to evaluate detector behavior under matched semantic content and to better separate generator-dependent cues from acoustic-class effects.

\begin{table}[t]
\centering
\scriptsize
\caption{Shared-prompt diagnostic analysis with EAT-AASIST.}
\label{tab:shared_prompt_diagnostic}
\begin{tabular}{lccc}
\hline
Model & Overall fake score & Var. mean & Range mean / median \\
\hline
Pre-trained & 0.5135 & 0.1413 & 0.8915 / 0.9461 \\
SynSFX fine-tuned & 0.9865 & 0.0082 & 0.0899 / 0.0037 \\
\hline
\end{tabular}
\end{table}

\section{Conclusion}
In this work, we introduced SynSFX, a multi-generator benchmark for isolated non-speech audio deepfake detection. Our experiments show that joint-domain training can mitigate speech-domain forgetting while maintaining strong performance on seen SynSFX generators. However, performance still degrades under unseen-generator evaluation, suggesting that current detectors remain sensitive to generator- and source-specific artifacts.

By releasing SynSFX with transparent provenance and a shared-prompt subset, we aim to support controlled studies of semantic content, generator effects, and cross-domain robustness. We also acknowledge several limitations: the dataset cannot cover the full diversity of real-world soundscapes, the unseen-generator setting remains limited in scale, and resolving cross-generator generalization is beyond the scope of this work. Future work will extend the benchmark and explore more robust representation learning methods.

\section{Data Availability}
The full dataset is available at \url{https://ofspectrum.com/news/synsfx}.

\section{Generative AI Use Disclosure}
Large Language Models (LLMs) were used solely for manuscript polishing (e.g., rephrasing and grammar checks) to improve clarity and readability. The LLMs were not used for ideation, methodology, experimental design, data analysis, or result interpretation. All scientific content was produced and verified by the authors.

\bibliographystyle{IEEEtran}
\bibliography{refs}

\end{document}